\documentclass[aps,pre,twocolumn,showkeys,superscriptaddress,longbibliography]{revtex4-1}
\usepackage{epsfig}
\usepackage{times}
\begin{document}

\title{Second-order freeriding on antisocial punishment restores the effectiveness of prosocial punishment}

\author{Attila Szolnoki}
\email{szolnoki.attila@energia.mta.hu}
\affiliation{Institute of Technical Physics and Materials Science, Centre for Energy Research, Hungarian Academy of Sciences, P.O. Box 49, H-1525 Budapest, Hungary}

\author{Matja{\v z} Perc}
\email{matjaz.perc@uni-mb.si}
\affiliation{Faculty of Natural Sciences and Mathematics, University of Maribor, Koro{\v s}ka cesta 160, SI-2000 Maribor, Slovenia}
\affiliation{CAMTP -- Center for Applied Mathematics and Theoretical Physics, University of Maribor, Mladinska 3, SI-2000 Maribor, Slovenia}
\affiliation{Complexity Science Hub, Josefst{\"a}dterstra{\ss}e 39, A-1080 Vienna, Austria}

\begin{abstract}
Economic experiments have shown that punishment can increase public goods game contributions over time. However, the effectiveness of punishment is challenged by second-order freeriding and antisocial punishment. The latter implies that non-cooperators punish cooperators, while the former implies unwillingness to shoulder the cost of punishment. Here we extend the theory of cooperation in the spatial public goods game by considering four competing strategies, which are traditional cooperators and defectors, as well as cooperators who punish defectors and defectors who punish cooperators. We show that if the synergistic effects are high enough to sustain cooperation based on network reciprocity alone, antisocial punishment does not deter public cooperation. Conversely, if synergistic effects are low and punishment is actively needed to sustain cooperation, antisocial punishment does act detrimental, but only if the cost-to-fine ratio is low. If the costs are relatively high, cooperation again dominates as a result of spatial pattern formation. Counterintuitively, defectors who do not punish cooperators, and are thus effectively second-order freeriding on antisocial punishment, form an active layer around punishing cooperators, which protects them against defectors that punish cooperators. A stable three-strategy phase that is sustained by the spontaneous emergence of cyclic dominance is also possible via the same route. The microscopic mechanism behind the reported evolutionary outcomes can be explained by the comparison of invasion rates that determine the stability of subsystem solutions. Our results reveal an unlikely evolutionary escape from adverse effects of antisocial punishment, and they provide a rationale for why second-order freeriding is not always an impediment to the evolutionary stability of punishment.
\end{abstract}


\maketitle

\section{Introduction}
Cooperation is widespread in human societies \cite{axelrod_84, henrich_aer01, fehr_tcs04, gachter_prsb10, rand_tcs13, capraro_pone14, hauser_n14}. Like no other species, we champion personal sacrifice for the common good \cite{hrdy_11, nowak_11}. Not only are people willing to incur costs to help unrelated others, economic experiments have shown that many are also willing to incur costs to punish those that do not cooperate \cite{yamagishi_jpsp86, fehr_aer00, gurerk_s06, rockenbach_n06, gaechter_s08, ule_s09, jordan_n16}. Unfortunately, cooperation is jeopardized by selfish incentives to freeride on the selfless contributions of others. What is more, individuals that abstain from punishing such freeriders are often called second-order freeriders for their failure to bear the additional costs of punishment \cite{fehr_n04, milinski_n08}. Several evolutionary models have been developed to study the effects of punishment on cooperation \cite{boyd_pnas03, fowler_pnas05, nakamaru_jtb06, hauert_s07, sigmund_tee07, boyd_s10, sasaki_pnas12, perc_njp12, jordan_jtb17}, and it has been pointed out that second-order freeriding is amongst the biggest impediments to the evolutionary stability of punishment \cite{panchanathan_n04, helbing_ploscb10, hilbe_srep12, chen_xj_njp14}.

In addition to second-order freeriding, the effectiveness of punishment is challenged by antisocial punishment. The fact that non-cooperators sometimes punish cooperators has been observed experimentally in different human societies \cite{ denant2007punishment, janssen2010lab, balafoutas_el4, wu_jj_pnas09, gachter2011limits, herrmann_s08}, and it has been shown theoretically that this antisocial punishment can prevent the successful coevolution of punishment and cooperation \cite{rand_jtb09, rand_jtb10}. In fact, if antisocial punishment is an option, prosocial punishment may no longer increase cooperation, deteriorating instead to a self-interested tool for protecting oneself against potential competitors \cite{rand_nc11}. While the punishment of freeriders is aimed at increasing cooperation, antisocial punishment can be a form of retaliation for punishment received in repeated games \cite{cinyabuguma2006can, denant2007punishment}, or is simply aimed at cooperators without a retaliatory motive \cite{gachter2011limits, herrmann_s08}.

Given the potential drawbacks associated with punishment related to second-order freeriding and antisocial punishment, it has been rightfully pointed out that the maintenance of cooperation may be better achievable through less destructive means. In particular, rewards may be as effective as punishment and lead to higher total earnings without potential damage to reputation or fear from retaliation \cite{dreber_n08, rand_s09}. Although many evolutionary models confirm the effectiveness of positive incentives for the promotion of cooperation \cite{hilbe_prsb10, hauert_jtb10, szolnoki_epl10, sasaki_jtb11, szolnoki_njp12, chen_xj_jrsi14, sasaki_bl14, okada_ploscb15, sasaki_srep15}, in this case too the challenges associated with second-order freeriding and antisocial rewarding persist \cite{dos-santos_m_prsb15, szolnoki_prsb15}.

Here we use methods of statistical physics to show how the two long-standing problems -- namely second-order freeriding and antisocial punishment -- cancel each other out in an unlikely and counterintuitive evolutionary outcome, and in doing so restore the effectiveness of prosocial punishment to promote cooperation. We extend the theory of cooperation by considering the spatial public goods game with non-punishing cooperators and defectors, as well as with cooperators who punish defectors and defectors who punish cooperators. As we will show in detail, spatial pattern formation leads to unconditional defectors forming an active layer around punishing cooperators, which protects them against defectors that punish cooperators. This is a new evolutionary escape from adverse effects of antisocial punishment, which in turn also reveals unexpected benefits stemming from second-order freeriding.

In what follows, we first present the spatial public goods game with prosocial and antisocial punishment, and then proceed with the results and a discussions of their implications for the successful coevolution of cooperation and punishment.

\section{Public goods game with prosocial and antisocial punishment}
The traditional version of the public goods game is simple and intuitive, and it captures the essence of the puzzle that is human cooperation \cite{sigmund_10,perc_pr17}. In a group of players, each one can decide whether to cooperate ($C$) or defect ($D$). Cooperators contribute a fixed amount (equal to $1$ without loss of generality) to the common pool, while defectors contribute nothing. The sum of all contributions is multiplied by a multiplication factor $r>1$, which takes into account synergistic effects of cooperation, and the resulting amount of public goods is divided equally amongst all group members irrespective of their strategies. Defection thus yields highest short-term individual payoffs, while cooperation is best for the group as a whole.

Here we extend this game by introducing two additional strategies, namely cooperators that punish defectors ($P_C$), and defectors that punish cooperators ($P_D$). The former represent prosocial punishment, while the later represent antisocial punishment. Technically, $P_C$ players punish $D$ and $P_D$ players, while $P_D$ players punish $C$ and $P_C$ players. In a $g$ group of size $G$ the resulting payoffs are
\begin{eqnarray}
\Pi_D^g &=& \Pi_{PGG} - \beta \frac{N_{P_C}}{G-1}\,\,,\\
\Pi_C^g &=& \Pi_{PGG} - \beta \frac{N_{P_D}}{G-1} - 1\,\,,\\
\Pi_{P_D}^g &=& \Pi_{PGG} - \beta \frac{N_{P_C}}{G-1} - \gamma \frac{N_C + N_{P_C}}{G-1}\,\,,\\
\Pi_{P_C}^g &=& \Pi_{PGG} - \beta \frac{N_{P_D}}{G-1} - \gamma \frac{N_D + N_{P_D}}{G-1} - 1\,\,,
\label{payoff}
\end{eqnarray}
where
\begin{equation}
\Pi_{PGG}^g = \frac{r (N_C + N_{P_C})}{G}
\end{equation}
and $N_C$, $N_D$, $N_{P_C}$ and $N_{P_D}$ are, respectively, the number of non-punishing cooperators, non-punishing defectors, punishing cooperators and punishing defectors in the $g$ group. In addition to the multiplication factor $r$ we have two additional parameters, which are $\beta$ as the maximal fine imposed on a player if all other players within the group punish her, and $\gamma$ as the maximal cost of punishment that can apply. Importantly, the values of both $\beta$ and $\gamma$ are kept the same for prosocial and antisocial punishment so as to not give either a default evolutionary advantage or disadvantage.

This public goods game is staged on a square lattice with periodic boundary conditions where $L^2$ players are arranged into overlapping groups of size $G=5$ such that everyone is connected to its $G-1$ nearest neighbors. Accordingly, each player belongs to $g=1,\ldots, G$ different groups, each of size $G$. Notably, the square lattice is the simplest of networks that takes into account the fact that the interactions among us are inherently structured rather than random. By using the square lattice, we continue a long-standing tradition that begun with the work of Nowak and May \cite{nowak_n92b}, and which has since emerged as a default setup to reveal all evolutionary outcomes that are feasible within a particular version of the public goods game \cite{perc_pr17}. We should note, however, that our observations are robust and do not restricted to this interaction topology. The only crucial criteria are players should have limited and stable connections with others, which allow network reciprocity to work.

Monte Carlo simulations are carried out as follows. Initially each player on site $x$ is designated either as a non-punishing cooperator, non-punishing defector, punishing cooperator or punishing defector with equal probability. The following elementary steps are then iterated repeatedly until a stationary solution is obtained, i.e., until the average fractions of strategies on the square lattice become time-independent. During an elementary step a randomly selected player $x$ plays the public goods game in all the $G$ groups where she is member, whereby her overall payoff $\Pi_{s_x}$ is thus the sum of all the payoffs $\Pi_{s_x}^{g}$ acquired in each individual group, as described above in Eqs.~(1-5). Next, a randomly selected neighbor of player $x$ acquires her payoff $\Pi_{s_y}$ in the same way.
Lastly, player $y$ imitates the strategy of player $x$ with a probability given by the Fermi function
\begin{equation}
\Gamma(s_x \to s_y)=1/\{1+\exp[(\Pi_{s_y}-\Pi_{s_x}) /K]\,\,,
\end{equation}
where $K$ quantifies the uncertainty by strategy adoptions \cite{szolnoki_pre09c}, implying that better performing players are readily adopted, although it is not impossible to adopt the strategy of a player performing worse. In the $K \to 0$ limit, player $y$ imitates the strategy of player $x$ if and only if $\Pi_{s_x} > \Pi_{s_y}$. Conversely, in the $K \to \infty$ limit, payoffs seize to matter and strategies change as per flip of a coin. Between these two extremes players with a higher payoff will be readily imitated, although the strategy of under-performing players may also be occasionally adopted, for example due to errors in the decision making, imperfect information, and external influences that may adversely affect the evaluation of an opponent. Without loss of generality we use $K=0.5$, in agreement with previous research that showed this to be a fully representative value \cite{szolnoki_pre09c, szolnoki_pre11, szolnoki_prx13}. Repeating all described elementary steps $L^2$ times constitutes one full Monte Carlo step (MCS), thus giving a chance to every player to change its strategy once on average.

We note that imitation is a fundamental process by means of which humans change their strategies \cite{colman_bbs03, cook_prsb12, naber_pnas13, buckert_jebo17}. The application of imitation-based strategy updating based on the Fermi function is thus appropriate and justified, although, as we will show, our results are robust to changes in the details that determine the microscopic dynamics of the studied public goods game. In terms of the application of the square lattice, we emphasize that, despite its simplicity, it fully captures the most relevant aspect of human interactions -- namely the fact that nobody interacts randomly with everybody else, not even in small groups, and that our interaction range is thus inherently limited. Applications of more complex interaction topologies are of course possible, but this does not affect our results. This is because our key argument is based on the limited number of interactions a players has, but it does not in any way rely on the specific properties of the square lattice topology.

The average fractions of all four strategies on the square lattice are determined in the stationary state after a sufficiently long relaxation time. Depending on the proximity to phase transition points and the typical size of emerging spatial patterns, the linear system size was varied from $L=400$ to $6000$, and the relaxation time was varied from $10^4$ to $10^6$ MCS to ensure that the statistical error is comparable with the size of symbols in the figures. We emphasize that the usage of a sufficiently large system size is a decisive factor that allows us to identify the correct evolutionary stable solutions. Using a too small system size may easily prevent this, for example if the linear size of the lattice is comparable to or smaller than the typical size of the emerging spatial patterns.

\section{Results}

Before presenting the main results, we briefly summarize the evolutionary outcomes in a well-mixed population. In the absence of a limited interaction range the behavior is largely trivial and resembles that reported before for the traditional two-strategy public goods game \cite{sigmund_10, helbing_ploscb10}. In particular, if $r$ exceeds the group size $G$ then both cooperative strategies dominate while all defectors die out. Conversely, below this threshold both defector strategies dominate while all cooperators die out. This behavior is also in agreement with the well-mixed results published in \cite{rand_jtb10}. In short, all the non-trivial evolutionary solutions  reported here in the continuation are due to the consideration of a structured population and remain completely hidden if a well-mixed population is assumed.

In what follows, we focus on two representative values of the multiplication factor that cover two relevantly different public goods game scenarios. First, we use $r=3.8$, where the spatial selection allows cooperators to survive even in the absence of punishment -- this is the well-known manifestation of network reciprocity, where the limited interactions among players allow cooperators to organize themselves into compact clusters, which confers them competitive payoffs in comparison to defectors \cite{nowak_06}. Subsequently, we also use a sufficiently small $r=3.0$ value, where cooperators can no longer survive solely due to network reciprocity and thus require additional support \cite{szolnoki_pre09c}. For both values of $r$ we determine the stationary fractions of strategies in dependence on the punishment fine $\beta$ and the punishment cost $\gamma$, and we pinpoint the location and type of phase transitions from the Monte Carlo simulation data.

\begin{figure}
\centerline{\epsfig{file=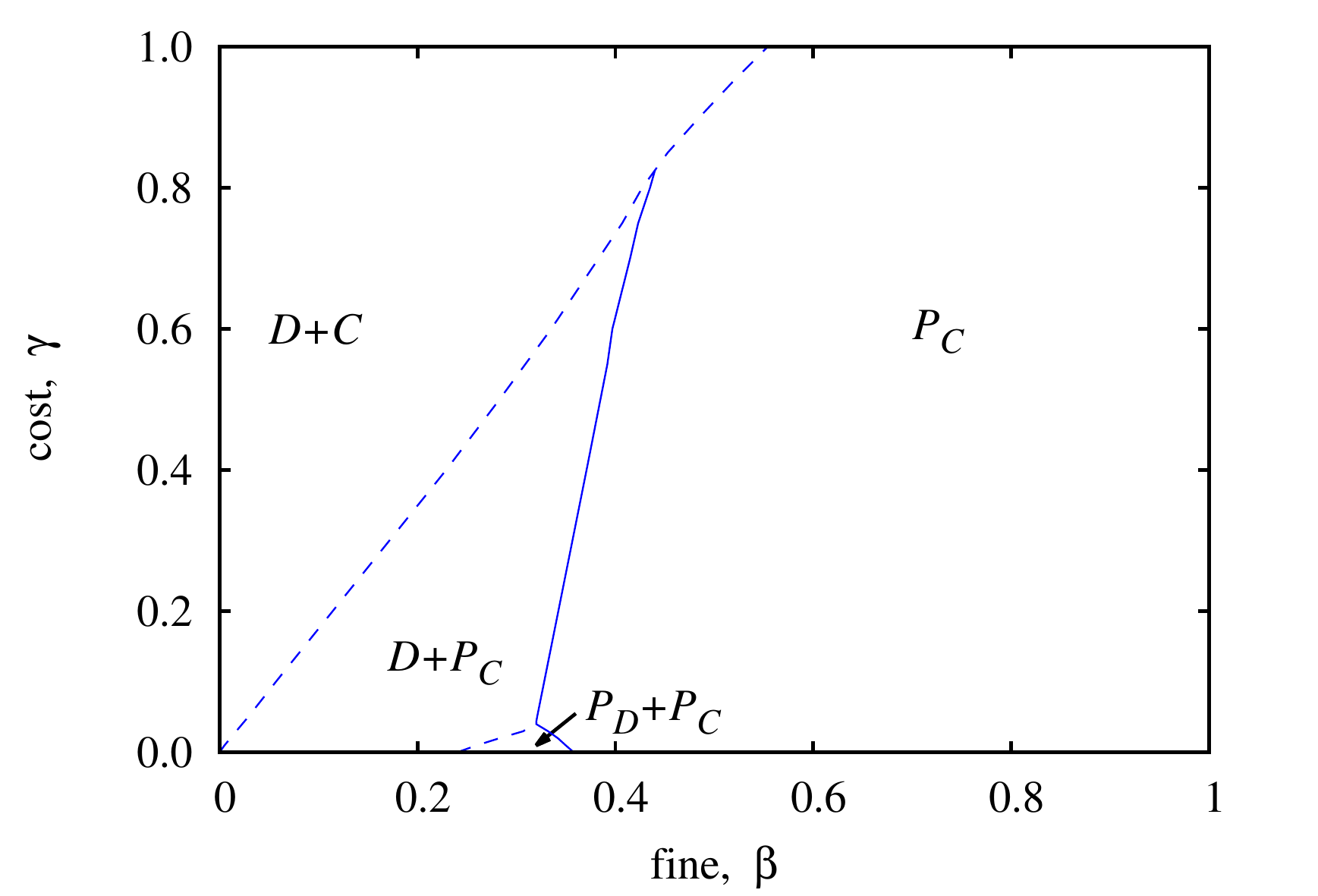,width=8.5cm}}
\caption{Full $\beta-\gamma$ phase diagram of the spatial public goods game with prosocial and antisocial punishment, as obtained for $r=3.8$. Solid lines denote continuous phase transitions while dashed lines denoted discontinuous phase transitions. Two representative cross-sections of this phase diagram are presented in Fig.~\ref{cross_r3_8}.}
\label{phd_r3_8}
\end{figure}

In Fig.~\ref{phd_r3_8}, we show the full $\beta-\gamma$ phase diagram, as obtained for $r=3.8$. As discussed above, we refer to this as the strong network reciprocity region. Presented results reveal that antisocial punishment is hardly viable, with $P_D$ players surviving only in a tiny region of the $\beta-\gamma$ parameter space. Conversely, as the fine $\beta$ increases, punishing cooperators subvert non-punishing cooperators, first via a discontinuous phase transition from the two-strategy $D+C$ phase to the two-strategy $D+P_C$ phase, and subsequently via a continuous phase transition to the absorbing $P_C$ phase. The discontinuous phase transition is due to indirect territorial competition, which emerges between $C$ and $P_C$ players competing against defectors \cite{helbing_ploscb10}, while the continuous phase transition is due to an increasing effectiveness of punishment that stems from the larger fines.

\begin{figure}
\centerline{\epsfig{file=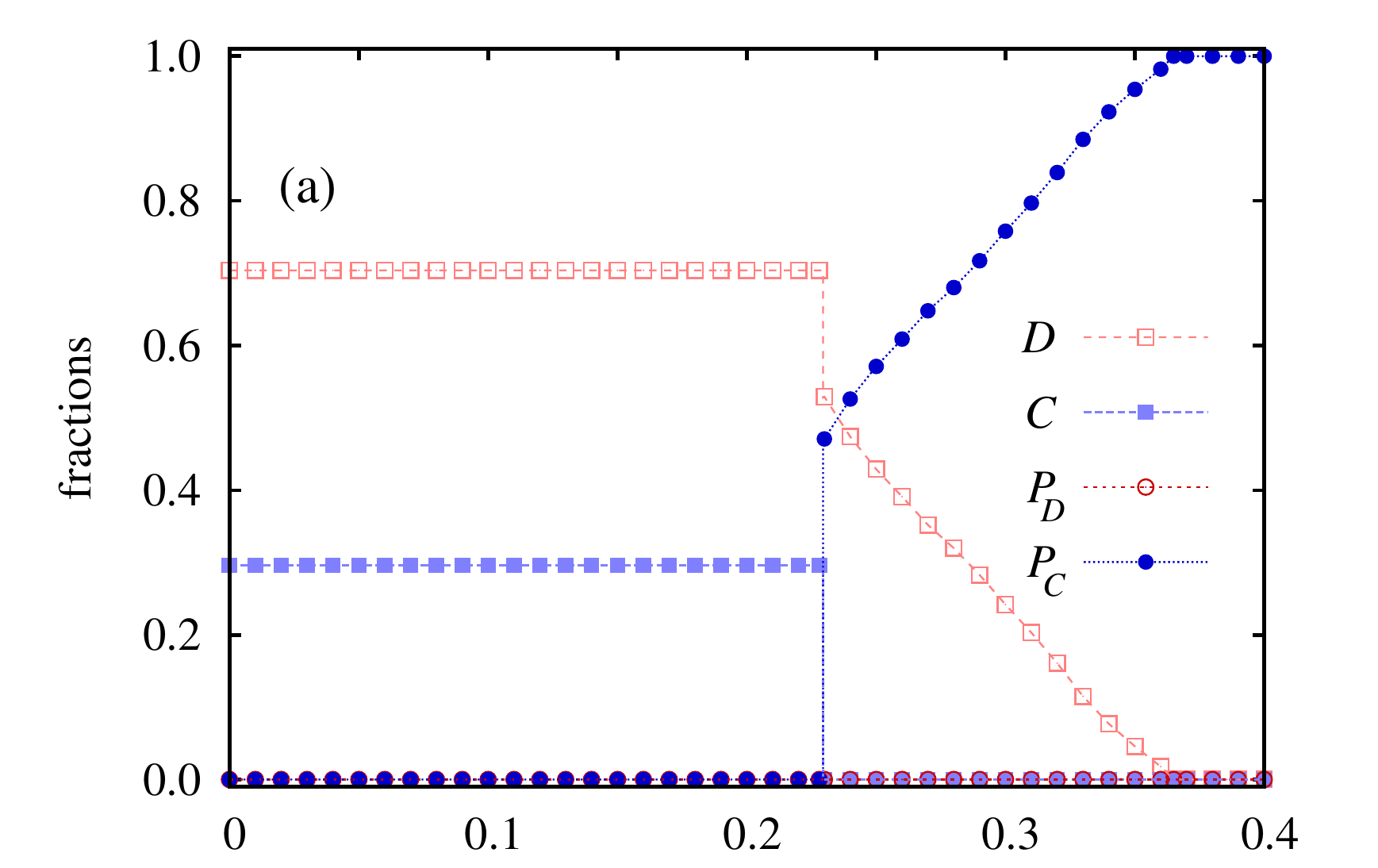,width=8.5cm}}
\centerline{\epsfig{file=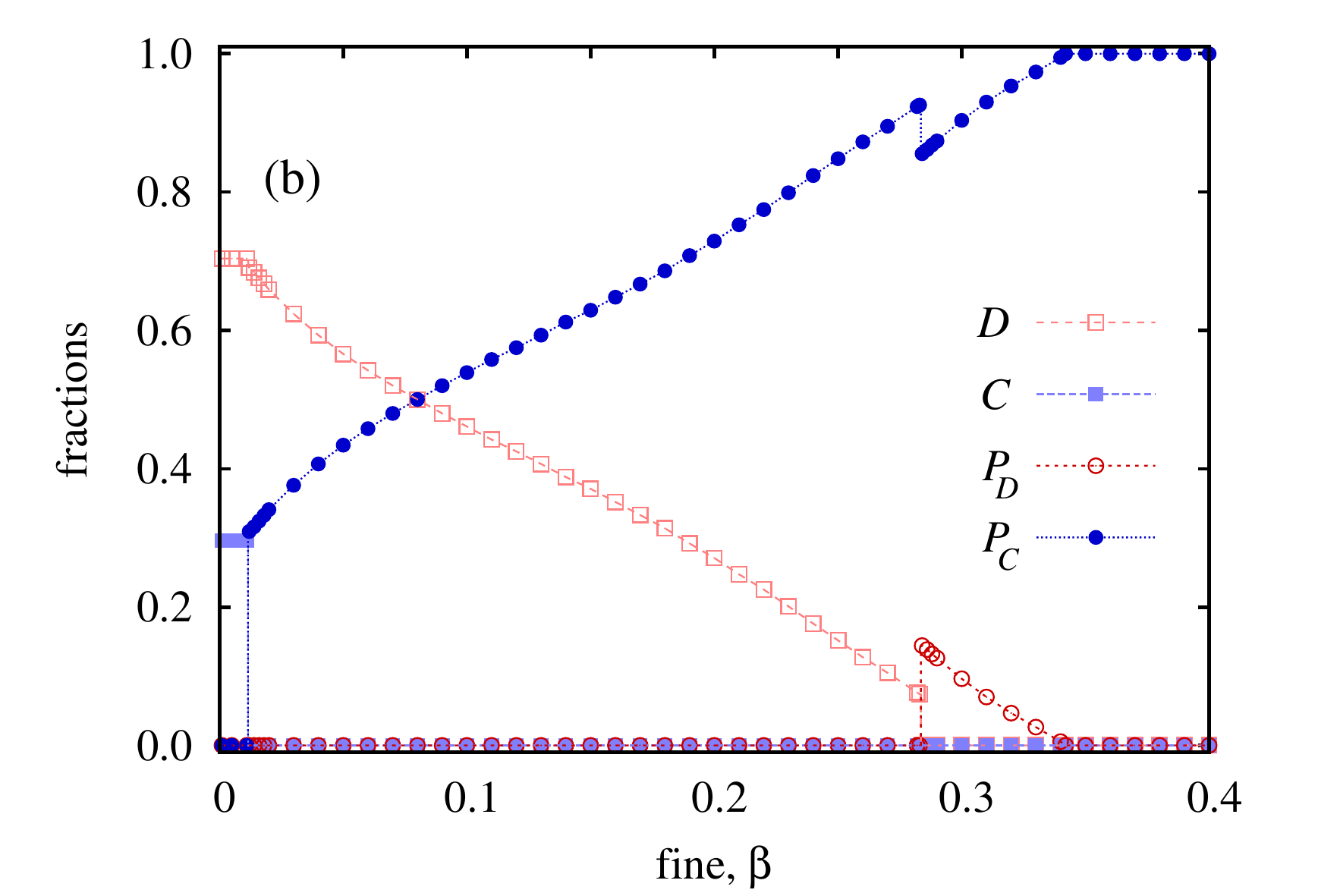,width=8.5cm}}
\caption{Two representative cross-sections of the phase diagram depicted in Fig.~\ref{phd_r3_8}, as obtained for the punishment cost $\gamma = 0.4$ (top) and $\gamma = 0.02$ (bottom). Depicted are stationary fractions of the four competing strategies in dependence on the punishment fine $\beta$.}
\label{cross_r3_8}
\end{figure}

The two representative cross-sections of the phase diagram in Fig.~\ref{cross_r3_8} provide a more quantitative insight into the nature of these phase transitions. In both cases the application of small punishment fines yields a punishment-free state, where traditional cooperators and defectors coexist due to network reciprocity. If the cost of punishment is considerable, as in panel~(a), the $D+C$ phase suddenly gives way to the $D+P_C$ phase at a critical value of the punishment fine $\beta_c=0.229$, and by increasing $\beta$ further, a defector-free state is reached at $\beta_c=0.361$. This succession of phase transitions remains the same if the cost of punishment is tiny, shown in panel~(b), apart from a narrow intermediate region of $\beta$, where antisocial punishment replaces non-punishing defectors via a discontinuous $D+P_C \to P_D+P_C$ phase transition at $\beta_c=0.284$. Interestingly, this phase transition is qualitatively identical to the preceding $D+C \to D+P_C$ phase transition -- in both cases non-punishing strategies are subverted by their punishing counterparts on the grounds of increasing punishment fines. It can also be observed that the emergence of a stable $P_D+P_C$ phase involves a slight decay of the fraction of $P_C$ players, although they quickly recover to full dominance as the punishment fine is increased further.

To sum up, in the strong network reciprocity region antisocial punishment has a negligible impact on the evolution of cooperation. In a small region of the $\beta-\gamma$ parameter space antisocial punishers can outperform non-punishing defectors to form a stable coexistence with prosocial punishers. But apart from this, and despite the fact that both forms of punishment are implemented equally effective (the values of both $\beta$ and $\gamma$ are kept the same for prosocial and antisocial punishment), antisocial punishment fails and is evolutionary uncompetitive.

\begin{figure}
\centerline{\epsfig{file=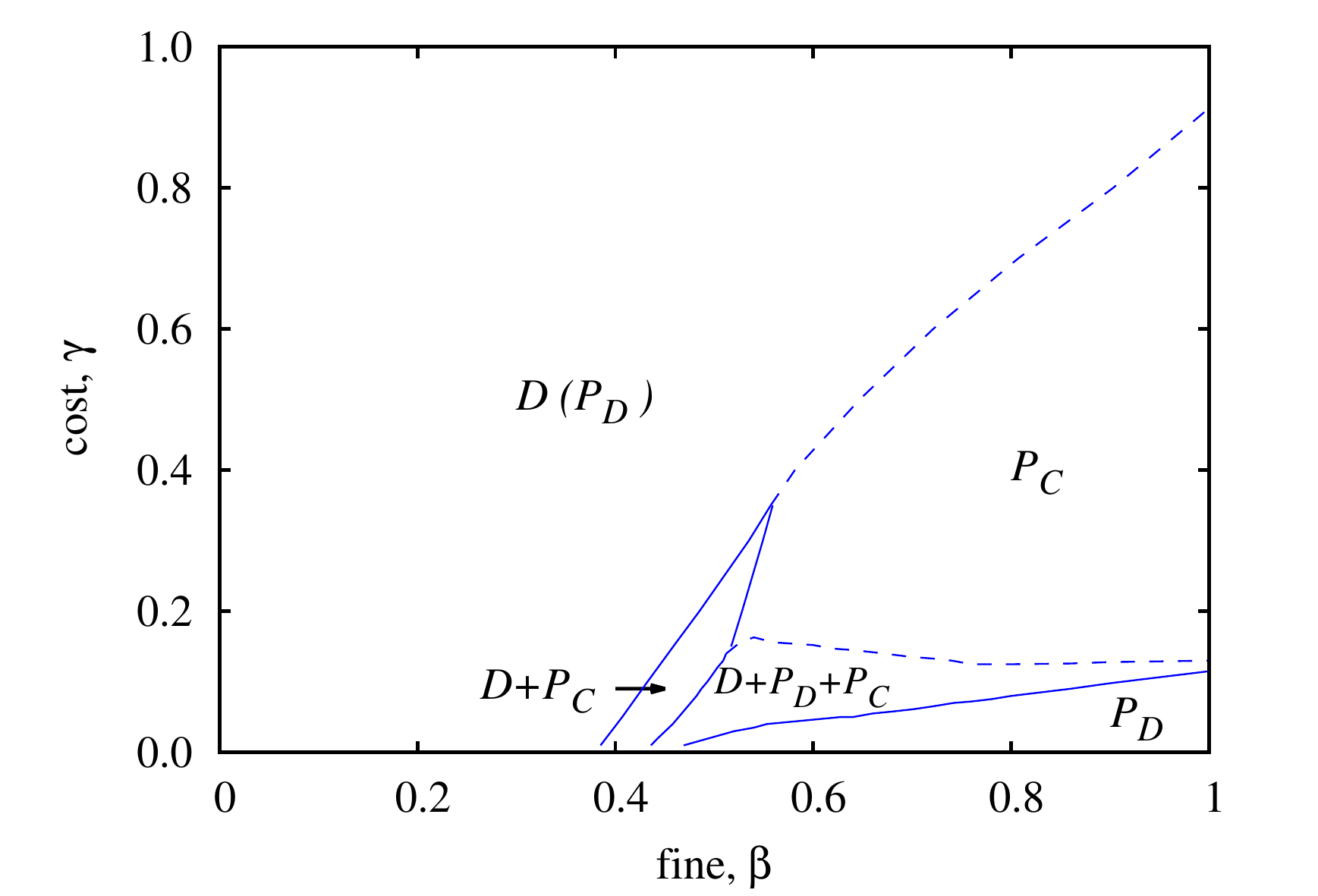,width=8.5cm}}
\caption{Full $\beta-\gamma$ phase diagram of the spatial public goods game with prosocial and antisocial punishment, as obtained for $r=3.0$. Solid lines denote continuous phase transitions while dashed lines denoted discontinuous phase transitions. Representative spatial evolutions of the four competing strategies are presented in Figs.~\ref{randominit}-\ref{to3cyc}, while two representative cross-sections of this phase diagram are presented in Fig.~\ref{cross_r3_0}.}
\label{phd_r3_0}
\end{figure}

An exciting question now is what if the network reciprocity alone is not strong enough to support the coexistence of cooperators and defectors? Although previous research has shown that prosocial punishment can be effective if the imposed fines are sufficiently high \cite{helbing_njp10,brandt_prsb03}, this results was obtained in the absence of antisocial punishment. However, if cooperators can also be punished the situation changes significantly. A subsystem analysis of the public goods game entailing only punishing cooperators and punishing defectors actually reveals that at $r=3.0$ cooperation is unable to survive regardless of the values of $\beta$ and $\gamma$, and regardless of the fact that punishing cooperators also benefit from network reciprocity.

Quite remarkably, the evolutionary outcome of the full 4-strategy public goods game can be very different. The full $\beta-\gamma$ phase diagram presented in Fig.~\ref{phd_r3_0} reveals that punishing cooperators can actually dominate completely in a sizable region of the parameter plane. Of course, if the cost of punishment is too high in relation to the imposed fines, defectors dominate. More precisely, $C$ and $P_C$ players die out fast, with only $D$ and $P_D$ players remaining. In the absence of the two cooperative strategies the relation between $D$ and $P_D$ players is neutral since the latter do not need to bear the punishment cost. This yields a logarithmically slow coarsening without surface tension, as in the voter model \cite{liggett_85, dornic_prl01}. In this case the probability to reach either the absorbing $D$ or the absorbing $P_D$ phase depends on the fraction of these two strategies \cite{cox_ap86} when all cooperators die out, which is typically higher for $D$ and hence the $D(P_D)$ notation in Fig.~\ref{phd_r3_0}.

\begin{figure}
\centerline{\epsfig{file=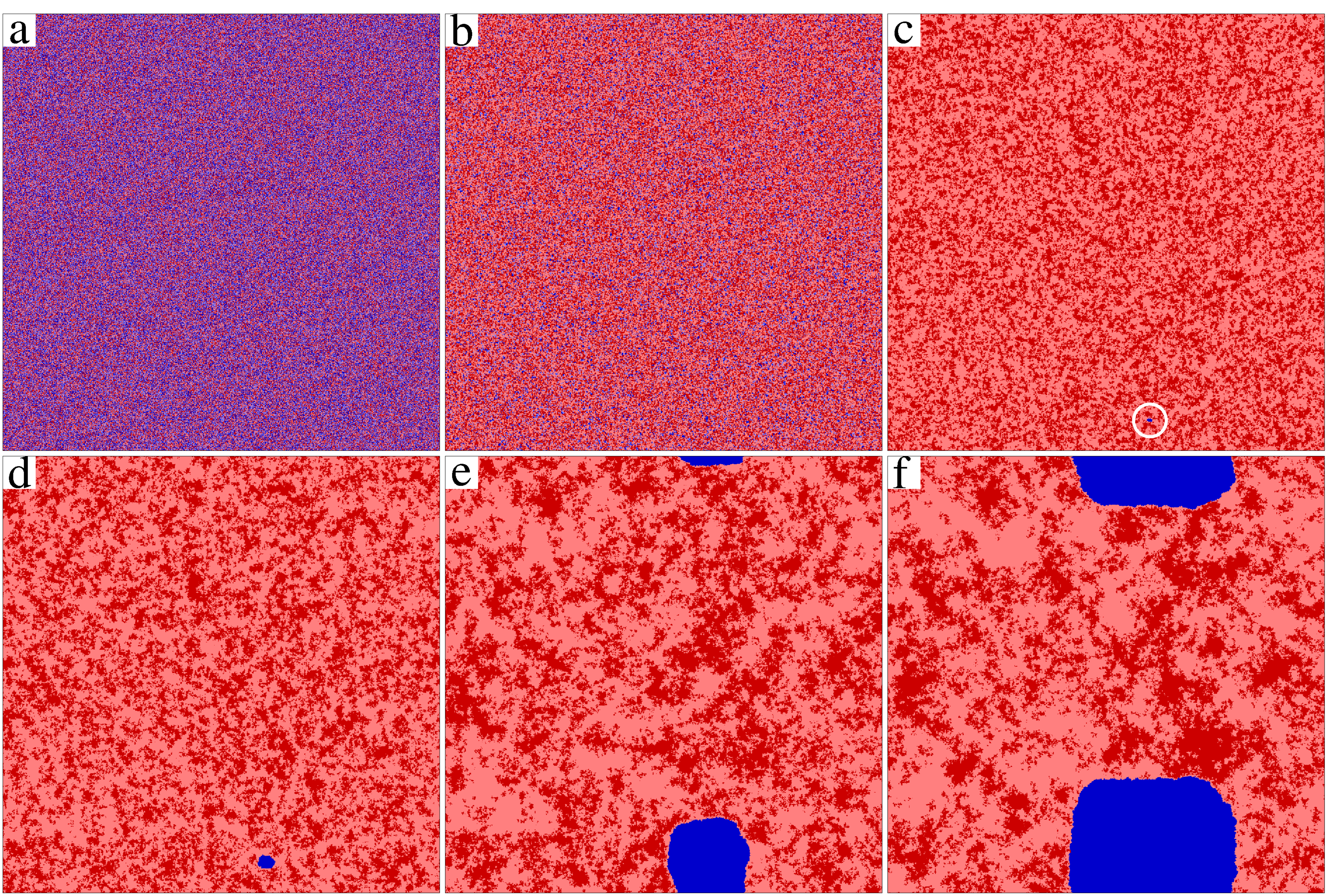,width=8.5cm}}
\caption{Representative spatial evolution of the four competing strategies. Depicted are snapshots of the square lattice, as obtained for $\beta = 0.8$, $\gamma = 0.36$, and $r=3$. Non-punishing (punishing) cooperators are depicted light blue (dark blue), while non-punishing (punishing) defectors are depicted light red (dark red). From a random initial state (a) both cooperative strategies start vanishing quickly (b). The only chance for cooperation to survive is if a lucky $P_C$ seed starts growing in the sea of $D$ players, as indicated in panel (c) by a white circle. It turns out, however, that $P_C$ players, surrounded by a thin active layer of $D$ players, can rise to complete dominance over time, as shown in panels (d-f) (the final state, where only $P_C$ remain after $6000$ MCS, is not shown). The linear size of the lattice is $L=800$.}
\label{randominit}
\end{figure}

Returning to the absorbing $P_C$ phase, since network reciprocity at $r=3.0$ is weak, an additional mechanism must be at work that allows the dominance of cooperation despite the low multiplication factor and despite antisocial punishment. This mechanism is illustrated in Fig.~\ref{randominit}, where we show a representative spatial evolution of the four competing strategies from a random initial state for parameter values that yield the absorbing $P_C$ phase. It is important to emphasize that a sufficiently large square lattice must be used, since otherwise the evolutionary process can quickly lead to a misleading outcome, i.e., to a solution that is not stable in the large population size limit, or to a solution that is highly sensitive on the initial fraction of strategies, as reported in \cite{rand_jtb10}. This is, however, just a finite-size effect because the evolutionary stable solution can spread in the whole population if it has a chance to emerge somewhere locally. Due to the random initial state the number of non-punishing and punishing cooperators starts dropping fast because support from network reciprocity is lacking, both because the value of $r$ is low, and even more so because compact clusters are not yet formed so early in the process. During this stage, if the population would be small, an accidental extinction of $P_C$ players would be very likely. Indeed, even with $L=800$ they manage to just barely survive, as indicated in panel (c) by a white circle. At this point the temporary winners appear to be $D$ and $P_D$ players, which in the absence of cooperators are neutral, and hence perform a logarithmically slow coarsening \cite{dornic_prl01}.

However, the unlikely evolutionary twist is yet to come and reveals itself in panels (d-f). Since $P_C$ players are weaker than $P_D$ players, the only chance for the former to survive is if they form a compact cluster inside a $D$ domain ($C$ can not survive either way because $r=3.0$ is too small). Although one might suspect that this ``hanging by a thread''-like survival of $P_C$ players is merely temporary because the superior $P_D$ players will eventually invade their cluster, this does in fact never happen. On the contrary, punishing cooperators eventually rise to complete dominance (the final state is not shown in Fig.~\ref{randominit}).

\begin{figure}
\centerline{\epsfig{file=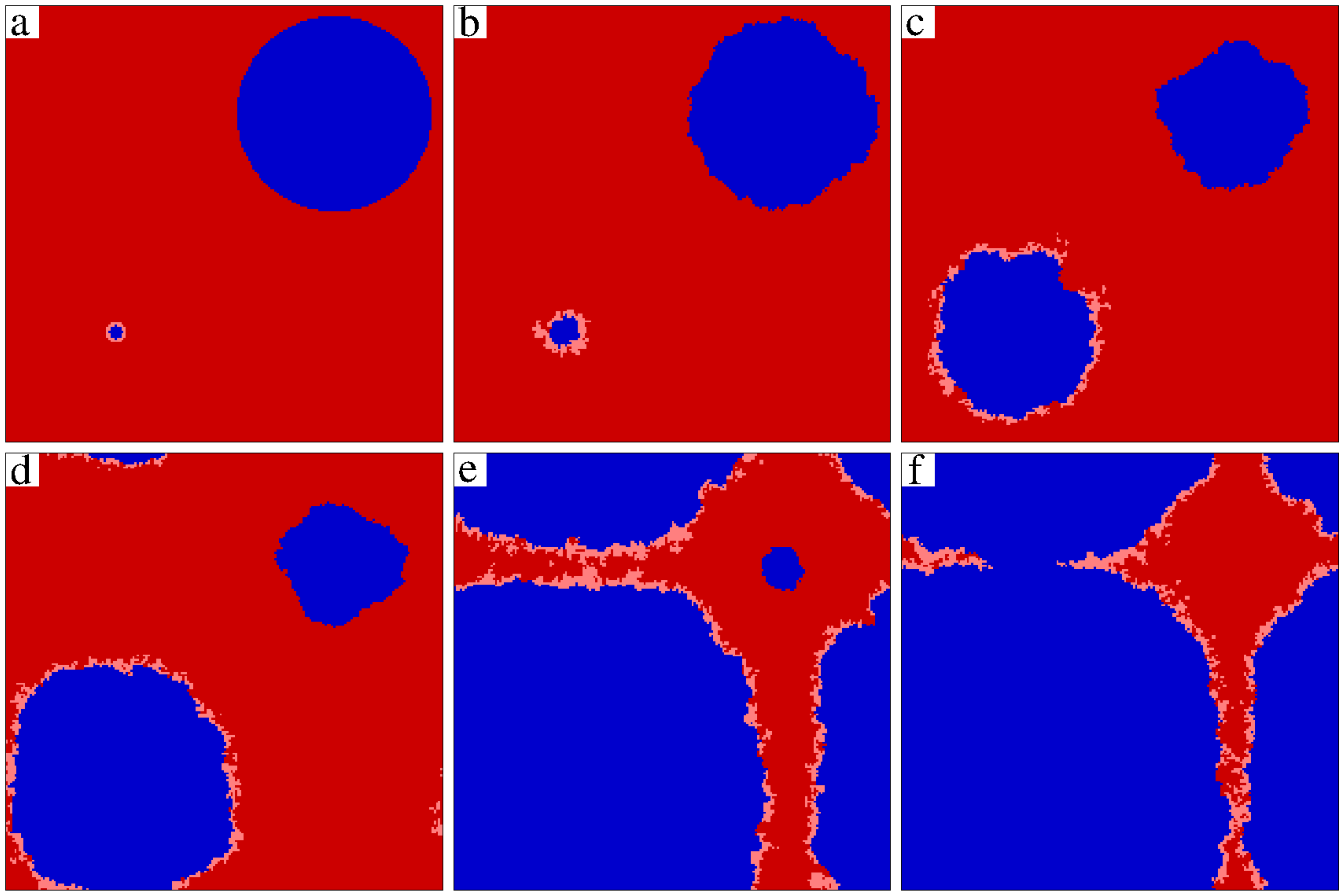,width=8.6cm}}
\caption{An illustration of how second-order freeriding on antisocial punishment restores the effectiveness of prosocial punishment. A small domain of $P_C$ players (dark blue), surrounded by a thin layer of $D$ players (light red) is inserted into the sea of $P_D$ players (dark red) in the bottom left corner of the lattice (a). Similarly, a sizable domain of $P_C$ players, but without the protective $D$ layer, is inserted into the sea of $P_D$ players in upper right corner of the lattice (a). While the large $P_C$ domain without the protective layer shrinks over time, the small $P_C$ domain with the protective $D$ layer grows (b-f). The absorbing $P_C$ phase is reached after $2000$ MCS (not shown). The linear size of the square lattice in this case is $L=200$. Other parameters are $\beta=0.8, \gamma=0.3$, and $r=3$.}
\label{dlayer}
\end{figure}

\begin{figure*}
\centerline{\epsfig{file=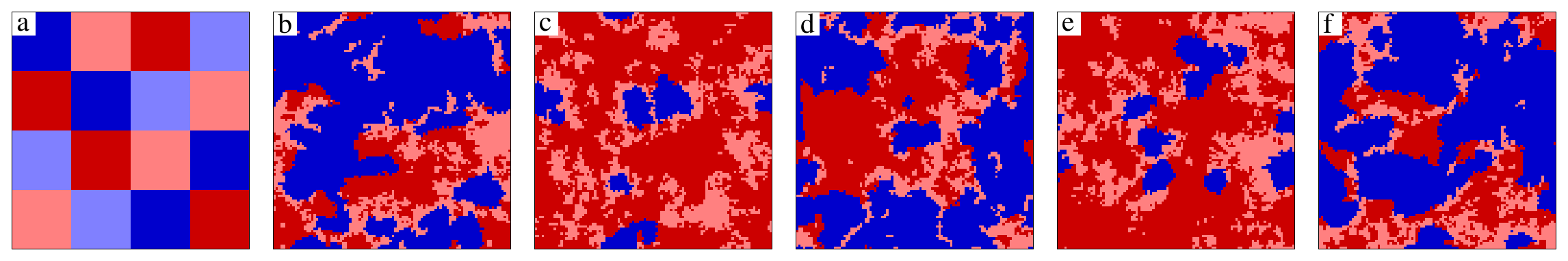,width=18cm}}
\caption{Representative spatial evolution of the four competing strategies from a prepared initial state towards the three-strategy $D+P_D+P_C$ phase that is sustained by cyclic dominance. Note that blue and red colors dominate cyclically over the course of $16000$ MCS from left to right. The colors used are the same as in Figs.~\ref{randominit} and \ref{dlayer}. Depicted are snapshots of the square lattice, as obtained for $\beta = 0.52$, $\gamma = 0.065$, and $r=3$. Since a prepared initial state is used, a small square lattice with linear size $L=100$ can be used for demonstration.}
\label{to3cyc}
\end{figure*}

Crucial for the understanding of this counterintuitive evolutionary outcome is the realization that punishing defectors suffer from second-order freeriding of non-punishing defectors as soon as they both meet in the vicinity of punishing cooperators. More precisely, when $D$ players meets with $P_D$ players in the vicinity of $P_C$ players, then $P_D$ players have to bear the additional cost of punishment while $D$ players are of course free from this burden. The same argument is traditionally put forward when it is time to explain why punishing cooperators are uncompetitive next to non-punishing cooperators near defectors, and why in fact punishment is evolutionary unstable. When antisocial punishment is present, however, this very same reasoning helps punishing cooperators to beat defectors that punish them. As a result, $D$ players start invading $P_D$ domains, but in parallel $P_C$ players also invade $D$ players from the other side of the interface. The thin active layer of $D$ players thus acts as a protection, shielding $P_C$ players from a direct invasion of $P_D$. As can be observed in panels (d-f), the shield is not passive, but expands permanently because $D$ players become successful when meeting $P_D$ players close to cooperators. (This process will be quantified via an effective invasion rate in the following section.) At the end, when $P_D$ players die out, the $D$ shield falls victim to the invasion of $P_C$ players, which thus rise to complete dominance. A direct illustration of this mechanism is shown in Fig.~\ref{dlayer}, where a prepared initial state was used for clarity. The comparison of the evolution of a large but lonely $P_C$ domain, and a tiny but $D$-protected seed of the same strategy illustrates nicely that the previously described ``activated layer" mechanism is effective to overcome the danger of simultaneous presence of antisocial punishment.

We show the above-described pattern formation in the animation provided in \cite{imit}. At this point, we also emphasize that the key mechanism that is responsible for the recovery of prosocial punishment is not restricted to the application of imitation-based strategy updating and is in fact robust to changes in the microscopic dynamics. For example, if we apply the so-called ``score-dependent viability'' strategy updating \cite{nakamaru_jtb97, rand_jtb10}, the trajectory of evolution remains the same \cite{score}. The only visible difference is that, in the latter case, the interfaces that separate the competing domains are rugged and strongly fluctuating, which in turn decelerates the evolutionary dynamics and prolongs the time needed to arrive at the same final outcome.

As we have pointed out, snapshots in Fig.~\ref{randominit} illustrate clearly that the size of the lattice plays a decisive role in reaching the correct evolutionary outcome from a random initial state. For some parameter values that bring the population closer to a phase transition point or because of the large fluctuations of strategy abundance during the pattern formation even $L=6000$ (linear system size) can turn out to be too small. In such cases a prepared initial state, as depicted in panel (a) of Fig.~\ref{to3cyc}, consisting of sizable patches of the four competing strategies, can help to determine the correct composition of the stationary solution. We have used this approach to determine the stability of the three-strategy $D+P_D+P_C$ phase, which according to the phase diagram in Fig.~\ref{phd_r3_0}, also forms an important part of the solution. Figure~\ref{to3cyc} illustrates that such a solution can be observed even if using a very small lattice size, if only suitable initial conditions are used. The alternating oscillations of read and blue indicate that this three-strategy phase is sustained by cyclic dominance. Indeed, due to using a different set of $\beta$ and $\gamma$ values from those used in Fig.~\ref{randominit}, here $P_D$ players beat $P_C$ players because of the low value of $r$, $P_C$ players beat $D$ players because of prosocial punishment, and $D$ players beat $P_D$ players near cooperators because of second-order freeriding. However, the balance of these invasions is such that neither strategy dies out, and hence the three-strategy $D+P_D+P_C$ phase is stable.

\begin{figure}[b]
\centerline{\epsfig{file=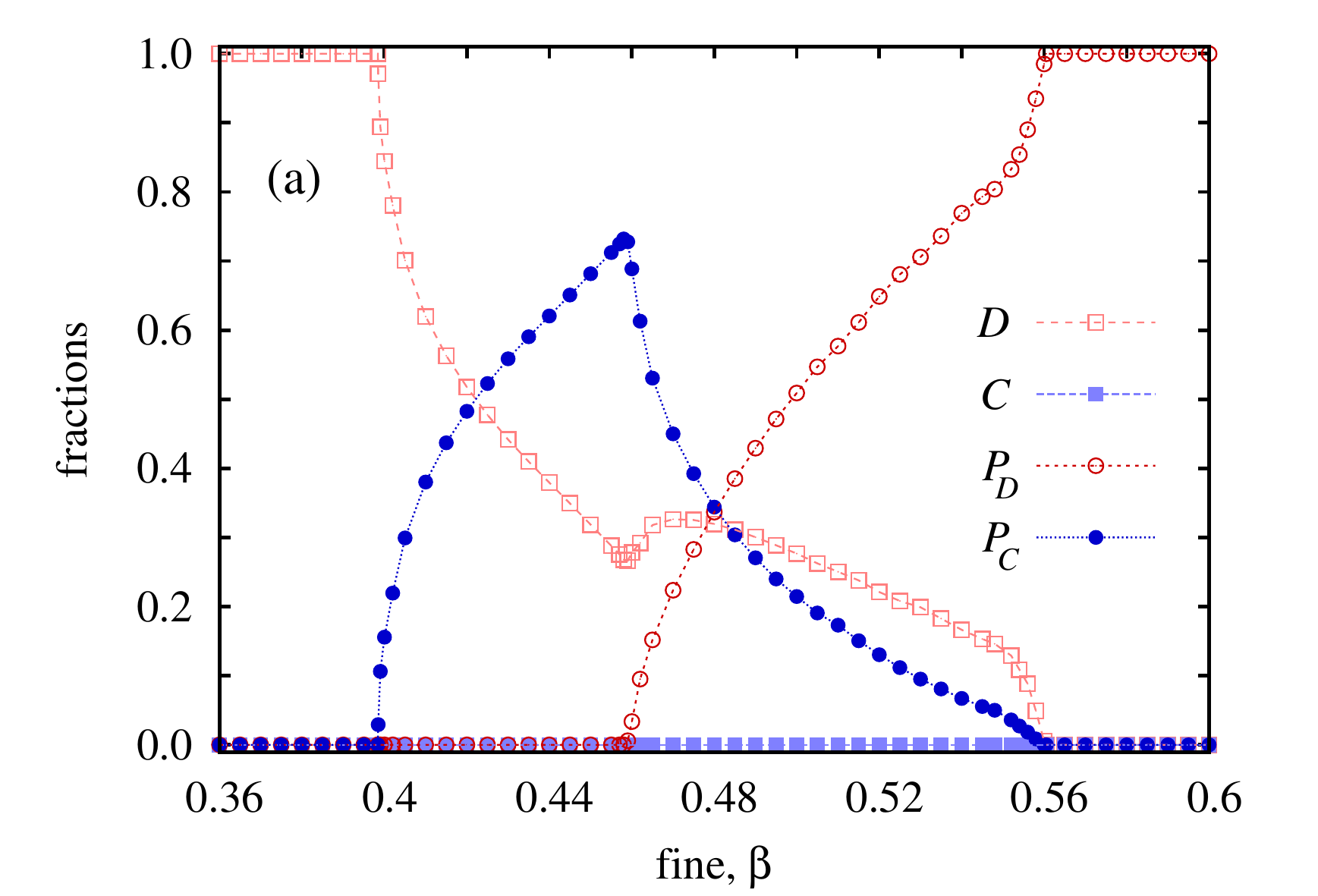,width=8.5cm}}
\centerline{\epsfig{file=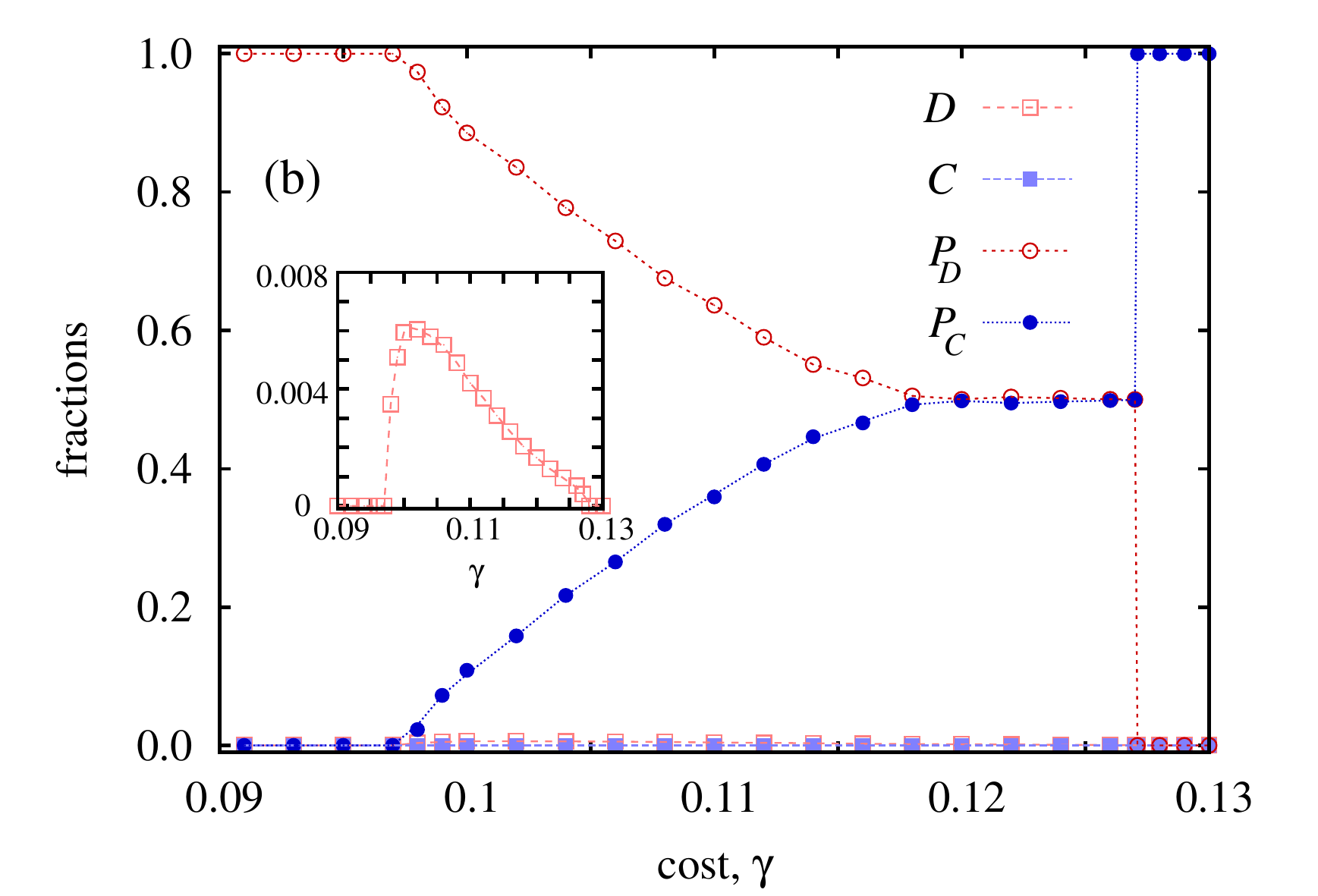,width=8.5cm}}
\caption{Two representative cross-sections of the phase diagram depicted in Fig.~\ref{cross_r3_0}, as obtained for the punishment cost $\gamma=0.04$ (top) and the punishment fine $\beta = 0.9$ (bottom). Depicted are stationary fractions of the four competing strategies in dependence on the other punishment parameter from the one used for the cross-section. The inset in the bottom panel shows just how tiny the fraction of non-punishing defectors in the stationary state can be.}
\label{cross_r3_0}
\end{figure}

The two representative cross-sections of the phase diagram in Fig.~\ref{cross_r3_0} show that the average fraction of competing strategies changes similarly as in the canonical rock-paper-scissors model \cite{szolnoki_jrsif14,reichenbach_n07, juul_pre12, intoy_pre15}. The stability of the three-strategy $D+P_D+P_C$ phase hinges  strongly on the continuous, albeit oscillating and sometimes nearly vanishing, presence of all three strategies. As the inset in the bottom panel of Fig.~\ref{cross_r3_0} shows, the average fraction of $D$ players can be extremely low, and therefore this three-strategy phase can be stable only if the size of the lattice is large enough. We have used $L=5400$ to produce results presented in Fig.~\ref{cross_r3_0}. If the size of the lattice would be smaller, a strategy could easily die out due to random fluctuations, in which case the evolution would terminate into a single-strategy phase.

\begin{figure}
\centerline{\epsfig{file=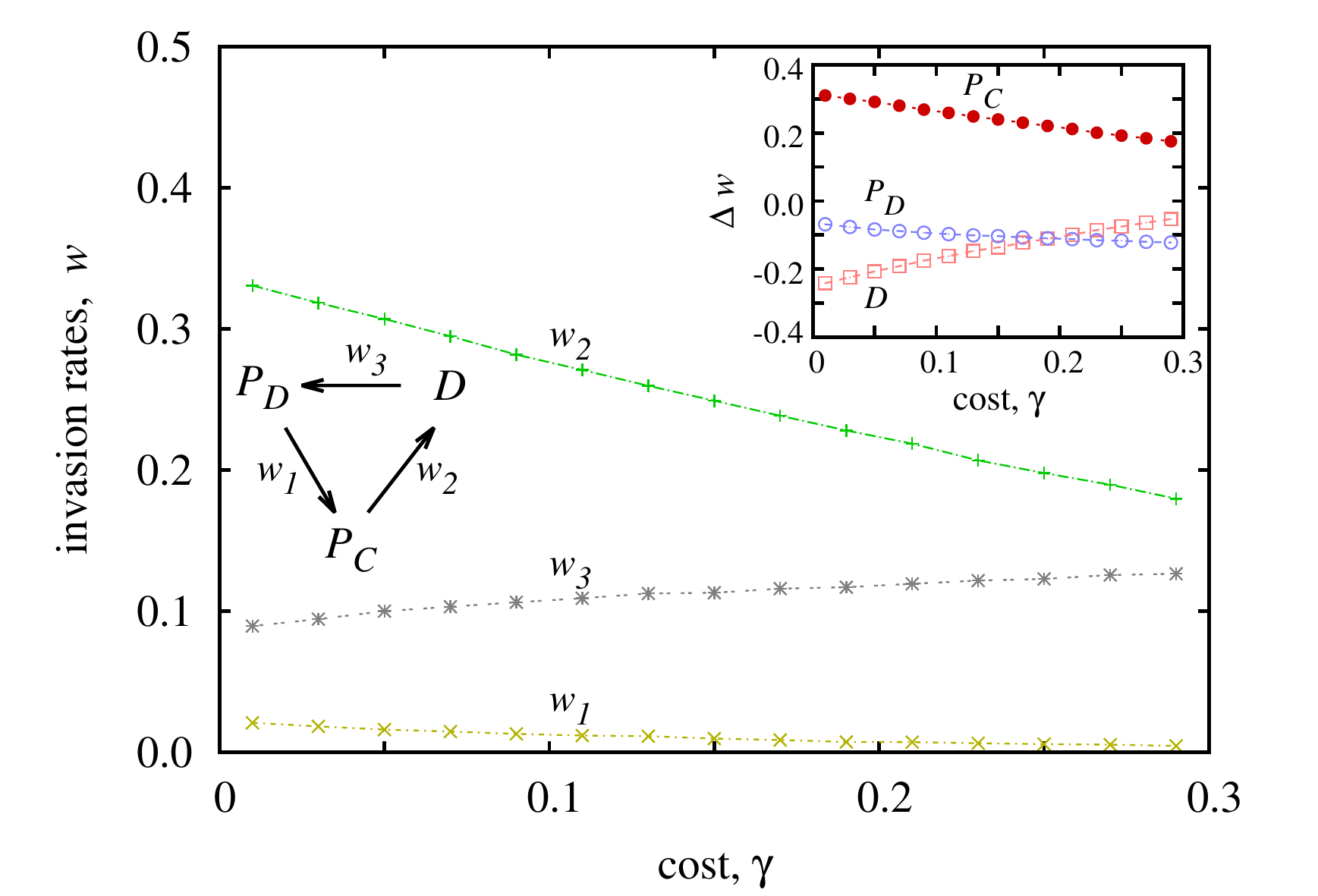,width=8.5cm}}
\caption{Invasion rates $w_i$ between three competing strategies (see inserted diagram) in dependence on the punishment cost $\gamma$, as obtained for $\beta = 0.8$ and $r=3.0$. We note that $D$ invades $P_D$ only in the vicinity of $P_C$ players, since otherwise the two defecting strategies are neutral. This fact is indicated by a dashed $w_3$ arrow. For details how this specific invasion rate is measured we refer to the main text. The inset shows the difference $\Delta w$ between different pairs of invasion rates, which explains how and why the fraction of strategies changes as a result of the increasing punishment cost (see main text for details).}
\label{inv}
\end{figure}

The invasion rates within the three-strategy $D+P_D+P_C$ phase, which exists under conditions that are more favorable for the survival of the $P_D$ strategy -- specifically, if the cost of punishment $\gamma$ is lower -- hence stabilizing the three-strategy phase, can be measured directly by monitoring the fractions of strategies when the evolution is initialized from straight domain interfaces \cite{szolnoki_pre14c, vukov_pre13}. While the meaning of $w_1$ and $w_2$ (see the diagram inserted in Fig.~\ref{inv}) is clear from the payoff differences, the determination of $w_3$ requires further clarification. Namely, if only $P_D$ and $D$ players would be present along the interface, then we would of course measure a net zero invasion rate because the two strategies are neutral in the absence of cooperators. However, since we are interested in their relation when $P_C$ players are present too, we use parallel interfaces of $P_C$ and $P_D$ players, separated by thin (width of $5$ lattice sites) layers of $D$ players. In this way, although $P_C$ and $P_D$ players do not interact directly, the setup properly describes the movement of $D$ layer that is followed by punishing cooperators. This ``effective'' invasion, which emerges only in the presence of the third party, is highlighted by a dashed arrow in the legend of Fig.~\ref{inv}.

The decay of $w_2$ in the main panel of Fig.~\ref{inv} highlights that the $P_C \to D$ invasions are relevant and in fact occur rather frequently, but also that their intensity deteriorates as the cost of punishment increases. Similarly, the $P_D \to P_C$ invasions are also a recurring phenomenon based on the positive value of the corresponding invasion rate $w_1$, which indicates that $P_D$ players would dominate $P_C$ players during a direct competition as a consequence of the small value of $r$. Nevertheless, this invasion rate also decays slightly as $\gamma$ increases because the costs associated with the main public goods game come to play second fiddle to the costs of punishment that both these strategies should bear. Lastly, the $w_3(\gamma)$ function is also always positive, because even a small cost evokes the second-order freeriding effect (in this case of course associated with avoiding the costs of antisocial punishment), such that in the presence of $P_C$ players $D$ players can invade $P_D$ players. Accordingly, as the value of $\gamma$ increases, so does the $w_3$ invasion rate, as shown in Fig.~\ref{inv}, which illustrates directly that second-order freeriding on antisocial punishment is responsible for the evolutionary success of prosocial punishment, specifically for the survival or even for the complete dominance of the $P_C$ strategy.

The differences between these three invasion rates, depicted in the inset of Fig.~\ref{inv}, allow us to understand how the relative abundance of strategies changes as a result. For example, $w_3 - w_2$ quantifies how the fraction of non-punishing defectors changes when we vary the cost of punishment. An increase in the value of $\gamma$ will support strategy $D$, but the actual beneficiary will be her predator, which is strategy $P_C$. This seemingly paradoxical response of the population to the increase of the punishment cost is a well-known consequence of cyclic dominance, i.e., when directly supporting a particular species will actually support her predator \cite{frean_prsb01}. This in turn explains why $P_C$ players rise to full dominance when we increase the value of $\gamma$, as well as why $P_D$ players dominate when we decrease it. Namely, decreasing the punishment cost supports the $P_C$ strategy, which is the prey of punishing defectors.

\section{Discussion}
We have shown how second-order freeriding on antisocial punishment restores the effectiveness of prosocial punishment, thus providing an unlikely and counterintuitive evolutionary escape from adverse effects of antisocial punishment. When the synergistic effects of cooperation are low to the point of network reciprocity failing to sustain it, cooperators that punish defectors can still rise to dominance because non-punishing defectors enable their evolutionary success by capitalizing on second-order freeriding and eliminating antisocial punishers as a result. If conditions for punishment are somewhat more lenient, we have shown that a three-strategy phase consisting of non-punishing defectors, punishing cooperators, and punishing defectors becomes stable. The relations within this phase, and its termination to an absorbing punishing cooperator or an absorbing punishing defector phase, can be fully understood in terms of invasion rates along straight interfaces that separate different strategy domains. We have demonstrated that these results are robust to changes in the microscopic dynamics, and we have emphasized that the only important property of the interaction structure is the limited interaction range rather than its topological details. Indeed, the mechanism relies solely on spatial pattern formation, and is the first stand-alone remedy against adverse effects of antisocial punishment, not relying on any additional strategic complexity or other assumptions limiting its general validity. Paradoxically, it turns out that antisocial punishment is vulnerable to the same second-order freeriding that is traditionally held responsible for preventing evolutionary stability of prosocial punishment.

We emphasize that these phenomena can not be observed in well-mixed populations. Furthermore, a reliable study of competing subsystem solutions requires a careful finite-size analysis of the spatial system. Additionally, the usage of random initial conditions may be misleading, especially if using a small system size, because it does not necessarily allow for all possible subsystem solutions to emerge (before they could compete with one another). This difficulties can be overcome by using suitable prepared initial states, which allow the evolutionary stable subsystem solution to form before competition between them unfolds. Since pattern formation and invasions of propagating fronts are general features of multi-strategy complex systems, such an analysis is a must when determining the consequences of spatiality.

Our research also reveals that under conditions that favor cooperation, for example when the multiplication factor of the public goods game is sufficiently high for the spatial selection alone to sustain cooperation, antisocial punishment is overall uncompetitive. In fact, even though we have used a fully symmetrical implementation of prosocial and antisocial punishment throughout our paper, antisocial punishers could survive only in very small regions of the parameter space. We may thus conclude that under such conditions cooperators that punish defectors should not be afraid of retaliatory antisocial punishment by defectors.

In comparison to previous findings concerning the symmetrical implementation of prosocial and antisocial rewarding \cite{szolnoki_prsb15}, we find that with punishment there is no lower bound on the multiplication factor that would be impossible to compensate with a sufficiently effective punishment system. But there is one with rewarding, i.e., below a critical value of the multiplication factor full defection is unavoidable, and this regardless of just how efficient the rewarding system might be. In case of punishment, ever lower values of the multiplication factor simply require ever higher fines at a given cost for cooperation to be sustained. Quite remarkably, the very same process puts a noose around antisocial punishers, which are defeated by second-order freeriding in their own ranks.

Although social preference models of economic decision-making predict that antisocial punishment should not occur \cite{fehr_qje99, rabin1993incorporating}, and despite the fact that antisocial punishment is also inconsistent with rational self-interest and the hypothesis that punishment facilitates cooperation, it is nevertheless remarkably common across human societies \cite{denant2007punishment, janssen2010lab, balafoutas_el4, wu_jj_pnas09, gachter2011limits, herrmann_s08}. In the light of this fact, it was important to extend the theory of cooperation in the spatial public goods game with the option that non-cooperators can punish cooperators. Rather unexpectedly, the detrimental effects of such antisocial punishment on the coevolution of punishment and cooperation turned out to be minor simply by taking into account the fact that the interactions among humans are inherently structured, entailing a limited number of frequently used links, rather than being random or well-mixed.

\begin{acknowledgments}
This research was supported by the Hungarian National Research Fund (Grant K-120785) and the Slovenian Research Agency (Grants J1-7009 and P5-0027).
\end{acknowledgments}

\end{document}